\documentclass{ws-ijmpb}

\begin{document}

\markboth{F. Benatti and R. Floreanini}
{Non-Positive Semigroup Dynamics in Continuous Variable Models}

%
\catchline{}{}{}{}{}
%

\title{NON-POSITIVE SEMIGROUP DYNAMICS IN CONTINUOUS VARIABLE MODELS}

\author{F. BENATTI}

\address{Dipartimento di Fisica Teorica, Universit\`a di Trieste,
\\
Strada Costiera 11, 34014 Trieste, Italy\\
and Istituto Nazionale di Fisica Nucleare, Sezione di Trieste\\
benatti@ts.infn.it}

\author{R. FLOREANINI}

\address{Istituto Nazionale di Fisica Nucleare, Sezione di Trieste,
\\
Dipartimento di Fisica Teorica, Universit\`a di Trieste,\\
Strada Costiera 11, 34014 Trieste, Italy\\
florean@ts.infn.it}

\maketitle


\begin{abstract}
Non-positive, Markovian semigroups are sometimes used to
describe the time evolution of subsystems immersed in an
external environment. A widely adopted prescription to avoid
the appearance of negative probabilities is to eliminate from
the admissible initial conditions those density matrices that would not
remain positive by the action of the semigroup dynamics.
Using a continuous variable model, we show that this procedure leads
to physical inconsistencies when two subsystems are considered and
their initial state is entangled.
\end{abstract}


\section{Introduction}

The time evolution of a subsytem in weak interaction with an external environment
can be modelled in terms of Markovian ({\it i.e.} memoryless) dynamical semigroups;
these are families $\{\gamma_t\}$ of one-parameter (=time) maps acting on the states
of the subsystem, typically described by density matrices, and obeying the forward in time
composition law, $\gamma_t\circ\gamma_s=\gamma_{t+s}$, $t,s\geq0$.
Such a description is very general and have been successfully used to describe various 
effects in open system dynamics, in quantum optics, 
quantum chemistry and atomic physics.\cite{1}$^-$\cite{12}

Nevertheless, the derivation of such time evolutions from the microscopic interaction
of the subsystem with the environment is often based on {\it ad hoc}
approximations, justified {\it a posteriori} with 
various physical considerations. As a result,
the obtained time evolutions might not be fully consistent:
in particular, such dynamics 
do not in general preserve the positivity
of the reduced density matrix.
Exceptions to this general result are obtained through the
use of rigorous mathematical treatments.\cite{1}$^-$\cite{5}

These inconsistencies are either dismissed as irrelevant for all
practical purposes\cite{13} or cured by adopting further 
ad hoc prescriptions.\cite{14}$^-$\cite{16}
In the latter case, the general attitude is to restrict the action of the
non-positive dynamical maps to a subset
of all possible initial reduced density matrices, those for which 
the time-evolution remains positive. This is equivalent to a suitable
selection of the initial conditions for the starting state of the subsystem,
a procedure sometimes referred to as ``slippage of the initial conditions''.
On physical grounds, this effect is viewed as the consequence of the
short-time correlations in the environment, that have not been
properly taken into account in the derivation of the Markovian limit
of the evolution, usually based on weak-coupling limit techniques
that get rid of transient effects.

In the following, we shall reexamine this widely used prescription to
cure possible inconsistencies produced by non-positive,
Markovian evolutions, and point out further potential problems
of this approach. Some results on finite-dimensional systems
have been reported in Ref.[17, 18]. In this note we shall instead
examine the case of simple infinite-dimensional systems.
More specifically, we shall study the behaviour of two
independent bosonic oscillators, one evolving with a 
dissipative Markovian semigroup, while the other
with a standard unitary dynamics. We shall limit our considerations
to the phenomenologically relevant set of Gaussian states\cite{19}$^-$\cite{27}
and to dissipative dynamics that preserve this set, the so-called
quasi-free semigroups.\cite{27}$^-$\cite{29}

Although very simple, this settings actually corresponds to a specific physical situation
in quantum optics\cite{30}$^-$\cite{33}, that of two coherent states travelling in
optical fibers, of which only one is subjected to noise,
producing dissipative phenomena,
while the other gives rise only to unitary, standard birefringence effects.
The considerations that follow may therefore have direct phenomenological
and experimental relevance.

We shall explicitly show that when the initial state
of the two oscillators is entangled, redefining the initial conditions
to make positive the single system time-evolution is not enough 
to cure all possible inconsistencies of the two-system dynamics.
Therefore, in order to have a physically acceptable
time evolution for the two subsystems in presence of entanglement, 
the above mentioned procedure of restricting initial conditions should 
be tested against the presence of non-classical correlations.

\section{Single Mode Dissipative Dynamics}

We shall first study the dynamics of a single oscillator
in weak interaction with an external environment, so that
its reduced time evolution can be well represented by
a Markovian semigroup.\cite{26} The states of this system
will be represented by a density matrix $\rho$, {\it i.e.}
by a positive Hermitian operator, with unit trace, acting
on a bosonic Hilbert space $\cal H$. It can be identified
with a Fock space, generated from the vacuum state through
the action of polynomials in the creation $a^\dagger$ and
annihilation $a$ operators; these operators obey the standard
bosonic oscillator algebra: $[a,a^\dagger]=1$, 
$[a^\dagger, a^\dagger]=[a,a]=0$.

Among all density matrices $\rho$, the so-called quasi-free or
Gaussian states are of particular interest\cite{19}$^-$\cite{27}: as mentioned before,
they can be easily produced in experiments in quantum optics.
These states are defined by the property that the expectations
$\langle W(z)\rangle={\rm tr}\big[W(z)\,\rho\big]$ of the Weyl operators
$W(z)=e^{za+z^* a^\dagger}$
are in Gaussian form, {\it i.e.} they are exponentials of a quadratic
form in the complex variable $z$ and its conjugate $z^*$.

Let us collect annihilation and creation operators in the column vector
$\bf a$ and its hermitian conjugate row vector 
${\bf a}^\dagger\equiv (a^\dagger\ a)$. Then Gaussian states 
are fully characterized 
by the $2\times 2$ matrix of bilinear correlations ($\mu,\nu=1,2$):\cite{23}$^-$\cite{27}
\begin{equation}
{\bf G}_{\mu\nu}\equiv\langle {\bf a}_\mu\, {\bf a}_\nu^\dagger\rangle
={\rm Tr}\big[{\bf a}_\mu\, {\bf a}_\nu^\dagger\, \rho\big]
=\left(\matrix{\langle a a^\dagger\rangle & \langle a a\rangle\cr
                          \langle a^\dagger a^\dagger\rangle & \langle a^\dagger a\rangle}\right)
\equiv\left(\matrix{\beta & \alpha\cr
                          \alpha^* & \beta-1}\right)\ .
\label{1}
\end{equation}
From its definition, it follows that the matrix $\bf G$ in (\ref{1}) is non-negative:
${\bf G}\geq0$, so that $\beta\geq1$ and $\beta(\beta-1)\geq |\alpha|^2$;
one can show that this is also a sufficient condition
for the state $\rho$ to represent a physical state.\cite{19}
For simplicity, we henceforth assume $\langle a^\dagger\rangle=\langle a \rangle=\,0$;
this condition can be easily released, by taking $\langle W(z)\rangle$
to include exponentials of linear terms in $z$ and $z^*$.

As mentioned in the introductory remarks, 
our analysis is based on the assumption that the time evolution of a single oscillator
immersed in a bath be Markovian and given by a semigroup, {\it i.e.}
by a trace preserving, one parameter
family of linear maps, acting on the set of density matrices $\rho$ representing the
oscillator states.
These maps are generated by master equations of the following 
Kossakowski-Lindblad form:\cite{1}$^-$\cite{5}
\begin{equation}
{\partial\rho(t)\over \partial t}=
-i\big[H ,\rho(t)\big] + L[\rho(t)]\ ,
\label{2}
\end{equation}
with $H$ an effective Hamiltonian; instead, $L$ is a linear map representing the dissipative piece:
it describes the effects of noise induced by the presence of the environment.
The semigroup evolutions that preserve the set of Gaussian states are generated
by equation in form (\ref{2}) with 
$H$ and $L$ that are quadratic in the creation and annihilation operators,\cite{27}$^-$\cite{29}
\begin{eqnarray}
\nonumber
&&H={1\over 2}\omega\, \{a^\dagger,\, a\}\ ,\qquad \omega\geq0\\
\nonumber
&&L[\rho]=\eta\Big(\big[a\rho, a^\dagger\big]+ \big[a,\rho a^\dagger\big]\Big)
+\sigma\Big(\big[a^\dagger\rho, a\big]+ \big[a^\dagger,\rho a\big]\Big)\\
&&\hskip 5cm -\lambda^*\big[a,[a,\rho]\big]
-\lambda\big[a^\dagger, [a^\dagger,\rho]\big]\ ;
\label{3}
\end{eqnarray}
the coefficients $\eta$, $\sigma$ and $\lambda$
encode the physical properties of the environment and can be expressed
in terms of the Fourier transform of the correlation functions
in the bath.\cite{1}$^-$\cite{5} 

By using the two-dimensional vectors
${\rm\bf a}$ and ${\rm\bf a}^\dagger$ introduced before,
the dissipative term in (\ref{3}) can be recast in the following compact form:
\begin{equation}
L[\rho]=\sum_{\mu,\nu=1}^{2} {\rm\bf C}_{\mu\nu} \bigg(
{\rm\bf a}_\nu\, \rho\, {\rm\bf a}_\mu^\dagger
-{1\over2} \Big\{ {\rm\bf a}_\mu^\dagger\, {\rm\bf a}_\nu\, ,\, \rho\Big\}
\bigg)\ ,
\label{4}
\end{equation}
where the bath coefficients are now embedded in the $2\times 2$
Kossakowski matrix:
\begin{equation}
{\rm\bf C}=\left(\matrix{\eta & \lambda^*\cr
                         \lambda & \sigma}\right)\ .
\label{5}
\end{equation}
The dissipative parameters in (\ref{5}) are not completely arbitrary. 
First of all, $\eta$ and $\sigma$ need to be
real in order to comply with the hermiticity preserving requirement 
of the generated semigroup. In addition,
the request that the finite-time evolution $\gamma_t$
generated by a master equation of the form (\ref{2}), (\ref{3}) 
map physical states into physical states, 
{\it i.e.} preserve the positivity of the density matrix $\rho(t)$
for all times, gives rise to further constraints.

The simplest way to guarantee the positivity of the dynamics $\gamma_t$
is to require the Kossakowski matrix in (\ref{5}) to be non-negative
\begin{equation}
{\rm\bf C}\geq0\ .
\label{6}
\end{equation}
Indeed, this condition is equivalent to the requirement of complete positivity, 
a stronger property than simple positivity,
that guarantees the physical consistency of the evolution $\gamma_t$ in all possible
situations.\cite{1}$^-$\cite{5} 
Dynamics of such type, obtained from the equation (\ref{2}), with Hamiltonian
as in (\ref{3}) and dissipative term as in (\ref{4}), (\ref{6}),
are known in the literature as quasi-free 
quantum dynamical semigroups:\cite{27}$^-$\cite{29}
as already observed, they are characterized by the property of transforming
the set of Gaussian states into itself.

Nevertheless, in phenomenological applications, the condition of complete positivity 
is often dismissed as unnecessary\cite{34}, since in many naive derivations
of the equation (\ref{2}) generating the reduced dynamics $\gamma_t$,
the hierarchy among the dissipative parameters imposed by
the condition (\ref{6}) appear not to be satisfied.\cite{13}$^-$\cite{16}
In such cases, it is easy to show that the resulting dynamics
$\gamma_t$ is not positive, hence physically unacceptable.

In order to explicitly show this, it is convenient
to consider the time evolution of the correlations in (\ref{1}),
as induced by the dynamics of the density matrix $\rho(t)$,
through the duality map: ${\rm Tr}\big[{\bf a}_\mu\, {\bf a}_\nu^\dagger\, \rho(t)\big]=
{\rm Tr}\big[({\bf a}_\mu\, {\bf a}_\nu^\dagger)(t)\, \rho\big]$.
One explicitly finds:\cite{27}$^-$\cite{29}
\begin{equation}
\partial_t {\bf G}(t)= {\bf A}^\dagger\cdot {\bf G}(t)
+{\bf G}(t)\cdot {\bf A}
+ {\bf B}\ ,
\label{7}
\end{equation}
where the $2 \times 2$ matrices ${\bf A}$ and ${\bf B}$ contain the
dependence on the hamiltonian $\omega$ and dissipative parameters
$\eta$, $\sigma$ and $\lambda$:
\begin{equation}
{\bf A}={1\over2}\left(\matrix{ \sigma - \eta +2i\omega & 0\cr
                            0 & \sigma-\eta- 2i\omega}\right)
\qquad
{\bf B}=\left(\matrix{ \eta & -\lambda^*\cr
                           -\lambda & \sigma}\right)\ .
\label{8}
\end{equation}
The positivity condition ${\bf G}(t)\geq0$ needs to be preserved for all times in order
for the the evolution $\gamma_t: \rho(0)\to\rho(t)$ to be positive
and thus physically acceptable.
In order to prove that this is not the case when (\ref{6}) is not satisfied,
{\it i.e.} when $\eta\sigma<|\lambda|^2$, it is sufficient to prove that
${\bf G}(t')<0$, for a certain $t'$. 
Let us assume that indeed this is the case; since
we start with a physical state, {\it i.e.} ${\bf G}(0)\geq0$, there will be a
time $t''<t'$ for which ${\bf G}(t'')$ has a null eigenvalue. Since we are dealing with 
evolutions of semigroup form, without loss
of generality, we can set $t''=\,0$. 

Let us now consider the average
${\bf Q}(t)\equiv\langle \Psi| {\bf G}(t) |\Psi\rangle$, where the two-component vector
$|\Psi\rangle$ is the eigenvector of ${\bf G}(0)$ corresponding to its
null eigenvalue. Recalling (\ref{1}), one easily finds
$\langle\Psi|=(-\alpha^*,\ \beta)$, with the condition $\beta(\beta-1)=|\alpha|^2$,
that assures the vanishing of the determinant of ${\bf G}(0)$. Clearly,
since ${\bf Q}(0)=\,0$,  ${\bf G}(t)$
will start acquiring a negative eigenvalue as soon as:
\begin{equation}
\partial_t {\bf Q}(0)<0\ .
\label{9}
\end{equation}
Using (\ref{7}) and (\ref{8}), this condition reduces to:
\begin{equation}
\eta(\beta-1) +\sigma\beta +2\,{\cal R}e(\alpha\lambda)<0\ .
\label{10}
\end{equation}
Since $\eta,\sigma\geq0$ and $\beta\geq1$, in order to satisfy (\ref{10})
one needs to choose the phase of $\alpha$ in such a way 
${\cal R}e(\alpha\lambda)$ be large and negative,
{\it i.e.} ${\cal R}e(\alpha\lambda)=-|\alpha|\ |\lambda|$.
Then, one can check that for $\eta>\sigma$, the condition
(\ref{10}) is always satisfied; this remains true also
in the case $\eta\leq\sigma$, provided $|\lambda|<\sigma$.
As a consequence, when the Kossakowski matrix in (\ref{5})
is negative, the evolution generated by (\ref{2}), (\ref{3})
results non-positive, and therefore physically unacceptable.

As mentioned in the Introduction, in order to cure this pathology
and continue to use semigroups of non-positive maps
to model reduced dynamics,
an {\it ad hoc} prescription has been proposed:\cite{14}$^-$\cite{16}
restrict the possible initial states $\rho(0)$ to those
for which $\rho(t)=\gamma_t[\rho(0)]$, $t>0$, as
generated by (\ref{2}), (\ref{3}), remains a 
physical Gaussian state. 
The argument is that any Markov approximation neglects a certain
initial span of time, namely a transient, during which no semigroup 
time-evolution is possible due to unavoidable memory effects.
The transient dynamics can be effectively depicted as a slippage
operator that, out of all possible initial states, selects those which
do not conflict with the Markov time-evolution when it sets in.
As we shall see in the next section, the slippage-argument may cure the
positivity-preserving problem, but appears to be inconclusive
when dealing with bi- and
multi-partite open quantum systems because of the existence of entangled
states.

\section{Two-Mode Dynamics and Entanglement}
 
We shall now extend the treatment discussed so far to the case
of two oscillators, one of which still immersed in
an external environment and therefore evolving
with the dissipative, semigroup dynamics generated by
(\ref{2}), (\ref{3}), while the other undergoes a unitary
evolution. As mentioned in the Introduction, such a system
can actually describe physically implementable situations in
quantum optics, {\it e.g.} that of polarized, coherent states
propagating along two fibers, one which is subjected to
noise, while the other is not. Having a physically consistent
theoretical treatment of the behavior of such setups is therefore
of great phenomenological relevance.

In this case of bipartite, two-mode settings, Gaussian states
$\rho$ are still characterized by having expectation
$\langle W(z_1, z_2)\rangle={\rm tr}\big[W(z_1, z_2)\,\rho\big]$ of the Weyl operators
$W(z_1, z_2)=e^{z_1 a_1 +z_2 a_2 + z^*_1 a_1^\dagger+z^*_2 a_2^\dagger}$
in Gaussian form. These states are again
fully characterized by the correlation matrix
${\bf G}$, which is now $4\times 4$, and still defined
as in (\ref{1}). The column vector $\bf a$ and its conjugate
${\bf a}^\dagger$ are now four-dimensional, with components
${\bf a}_\mu^\dagger=(a^\dagger_1, a_1, a^\dagger_2, a_2)$,
$\mu=1,2,3,4$, where $a_i^\dagger$, $a_i$, $i=1,2$, are
single mode creation and annihilation operators, obeying
the standard oscillator algebra:
$[a_i,\ a^\dagger_j]=\delta_{ij}$,
$[a_i^\dagger,\ a^\dagger_j]=\,0=[a_i,\ a_j]$.

The time evolution is again a Markovian semigroup
generated by a master equation of the form
(\ref{2}). The Hamiltonian is now the sum of two pieces:
\begin{equation}
H={1\over 2}\omega_1\, \{a_1^\dagger,\, a_1\}+
{1\over 2}\omega_2\, \{a_2^\dagger,\, a_2\}\ ,
\label{11}
\end{equation}
while the dissipative contribution can be written
as in (\ref{4}), with a $4\times 4$ Kossakowski matrix
with null entries, except in the upper left corner
where it coincides with (\ref{5}). Such compound dynamics ${\Gamma}_t=\gamma_t\otimes U_t$,
being represented by the tensor product of the dissipative
evolution $\gamma_t$ discussed in the previous section
for the first oscillator mode and of a unitary one $U_t$ for the second mode,
preserves the Gaussian form of the states $\rho$.

As in the single mode case, the time evolution equation can be
equivalently formulated in terms of the correlation matrix;
explicitly:
\begin{equation}
\partial_t {\bf G}(t)= {\cal A}^\dagger\cdot {\bf G}(t)
+{\bf G}(t)\cdot {\cal A}
+ {\cal B}\ ,
\label{12}
\end{equation}
where ${\cal A}$ is diagonal with entries
${1\over 2}(\sigma-\eta+ 2i\omega_1, \sigma-\eta- 2i\omega_1,
2i\omega_2, -2i\omega_2)$, while the inhomogeneous term is given by
\begin{equation}
{\cal B}=\left(\matrix{ {\bf B} & {\bf 0}\cr
                            {\bf 0} & {\bf 0} }\right)\ ,
\label{13}
\end{equation}
with $\bf 0$ representing the $2\times 2$ zero matrix and $\bf B$ is as in (\ref{8}).
The matrix $\bf G(t)$ can similarly be written in block form:
\begin{equation}
{\bf G}(t)=\left(\matrix{ {\bf G}_1(t) & {\bf W}(t)\cr
                            {\bf W}^\dagger(t) & {\bf G}_2(t) }\right)\ ,
\label{14}
\end{equation}
where ${\bf G}_1(t)$ and ${\bf G}_2(t)$ are $2\times 2$ correlation matrices
corresponding to the reduced single-mode Gaussian states
obtained by tracing over the degrees of freedom of the second,
${\rm Tr}_2[\rho(t)]$,
respectively first, ${\rm Tr}_1[\rho(t)]$, oscillator. As such, they must satisfy
the positivity condition: ${\bf G}_i(t)\geq0$, $i=1,2$, for all times.

Therefore, to avoid appearance
of negative probabilities when the single mode dynamics $\gamma_t$
is not positive, {\it i.e.} when $\eta\sigma<|\lambda|^2$, 
in line with the ``slippage'' prescription,
one has to restrict the choice of the initial condition $\rho(0)$
to those Gaussian density matrices for which the reduced states 
${\rm Tr}_2[\rho(0)]$ gives rise to positive correlation matrices,
${\bf G}_1(t)\geq0$. Although this prescription clearly works for
initial states in factorized form, it fails in general when $\rho(0)$
is entangled.

In order to make the incompatibility between slippage and
entanglement apparent, let us consider
an initial correlation matrix of the form:
\begin{equation}
{\bf G}(0)=\left(\matrix{ {\bf G}_1 & {\bf G}_1^{1/2} {\bf G}_2^{1/2}\cr
                          {\bf G}_2^{1/2} {\bf G}_1^{1/2} & {\bf G}_2 }\right)\ ,
\label{15}
\end{equation}
with the conditions ${\bf G}_i\geq 0$, $i=1,2$; note that the matrix in
(\ref{15}) satisfy the physical condition ${\bf G}(0)\geq0$ by construction.
Further, the $2\times 2$ correlation submatrices ${\bf G}_i$ 
are chosen in such a way that their positivity
is preserved under the action of the compound dynamics $\Gamma_t$.
Let ${\bf G}_2$ be invertible.
Then, the $4\times 4$ matrix in (\ref{15}) possesses a zero eigenvalue,
whose associated eigenvector can be written as
\begin{equation}
|\Psi\rangle=\left(\matrix{ |\psi\rangle\cr
\cr
-{\bf G}_2^{-1/2} {\bf G}_1^{1/2}|\psi\rangle }\right)\ ,
\label{16}
\end{equation}
where $|\psi\rangle$ is an arbitrary two-dimensional vector.
Let us then consider the average
${\cal Q}(t)\equiv \langle\Psi| {\bf G}(t) |\Psi\rangle$. Since
${\cal Q}(0)=\,0$, the matrix ${\bf G}(t)$ will start developing
negative eigenvalues as soon as
\begin{equation}
\partial_t {\cal Q}(0)<0\ .
\label{17}
\end{equation}
Explicitly, $\partial_t {\cal Q}(0)=\langle\Psi| {\cal B} |\Psi\rangle=
\langle\psi| {\bf B} |\psi\rangle$; 
therefore, the condition (\ref{17}) can always be satisfied by choosing for
the two-vector $|\psi\rangle$ the negative eigenvector of $\bf B$.

The physical inconsistency of dynamics for which the Kossakowski matrix
in (\ref{5}) is not positive is now apparent: the slippage prescription
make them positive, but not completely positive; therefore, when the
subsystem is bipartite, one can always find an entangled Gaussian state
that is mapped out from the set of physical states by the compound dynamics
$\Gamma_t$.

That the initial state $\rho(0)$ giving rise to the correlations in (\ref{15}) is entangled
can be directly verified using the operation of 
partial transposition\cite{35,36}, {\it i.e.}
of transposition involving only one of the two oscillators, say the second one.
It can be most easily implemented on symmetric correlation matrices:
\begin{equation}
{\rm\bf V}_{\mu\nu}={1\over 2} \Big\langle \big\{{\rm\bf a}_\mu\, ,
\ {\rm\bf a}_\nu^\dagger \big\}\Big\rangle\equiv
{1\over2}{\rm Tr}\Big[\big({\rm\bf a}_\mu {\rm\bf a}_\nu^\dagger
+{\rm\bf a}_\nu^\dagger {\rm\bf a}_\mu\big)\, \rho\Big]\ .
\label{18}
\end{equation}
where it amounts to the exchange $a_2 \leftrightarrow a_2^\dagger$. 
This transformation clearly maps states into states for separable ones,
but not in general for correlated ones: it thus provides
a sufficient criterion for bipartite entanglement,
which turns out to be also sufficient for two-mode
states, as in in the present case.\cite{37}

For the case at hand, one finds the the operation of partial
transposition with respect to the second system results
in the following transformation of the $4\times 4$ symmetric
covariance:
\begin{equation}
{\rm\bf V}\rightarrow \widetilde{\rm\bf V}=
{\rm\bf T}\cdot{\rm\bf V}\cdot{\rm\bf T}\ ,
\qquad
{\rm\bf T}=\left(\matrix{ {\bf 1} & {\bf 0} \cr
                          {\bf 0} & \sigma_1 }\right)\ ,
\label{19}
\end{equation}
where ${\bf 1}$ is the $2\times 2$ unit matrix, while
$\sigma_1$ is the first Pauli matrix.
Equivalently, for ${\bf G}(0)$ in {(\ref{15}), one finds:
\begin{equation}
{\bf G}(0)\rightarrow\widetilde {\bf G}(0)=
\left(\matrix{ {\bf G}_1 & {\bf G}_1^{1/2} {\bf G}_2^{1/2}\, \sigma_1\cr
                          \sigma_1\, {\bf G}_2^{1/2} {\bf G}_1^{1/2} & \widetilde{\bf G}_2 }\right)\ ,
\label{20}
\end{equation}
where $\widetilde{\bf G}_2$ indicates transposition of ${\bf G}_2$.
By construction, the initial density matrix $\rho(0)$
corresponds to a physical state, {\it i.e.}
${\bf G}(0)\geq0$;
if one further finds:
\begin{equation}
\widetilde{\bf G}(0)<0\ ,
\label {21}
\end{equation}
then the state $\rho(0)$ is surely entangled. One can show that (\ref{21})
is equivalent to the following condition involving $2\times 2$ submatrices
of ${\bf G}(0)$:\cite{38}
\begin{equation}
{\bf G}_1 -{\bf G}_1^{1/2}{\bf G}_2^{1/2}\, \big[\widetilde{\bf G}_2\big]^{-1}\,
{\bf G}_2^{1/2} {\bf G}_1^{1/2}<0\ ,
\label {22}
\end{equation}
which in turn reduces to:
\begin{equation}
\big[{\bf G}_2\big]^{-1}-\sigma_1\, \big[\widetilde{\bf G}_2\big]^{-1}\,
\sigma_1<0\ .
\label {23}
\end{equation}
Since ${\bf G}_2$ can always be written as in (\ref{1}), with
$\beta(\beta-1)>|\alpha|^2$, by direct computation, one immediately finds that the
combination in the l.h.s. of (\ref{23}) always possesses a negative eigenvalue,
and therefore that $\rho(0)$ is indeed an entangled state.

\section{Discussion}

The dynamics of a subsystem in weak interaction with an external environment
can be described in terms of semigroups of linear maps $\gamma_t$
generated by a Markovian master equation.
Although rigorously proven only in special cases, this result 
is nevertheless believed to hold in general, since memory 
effects should disappear from the reduced subsytem dynamics
as soon as all correlations in the environment have died out.

This mathematical effective description
of reduced dissipative dynamics needs to
satisfy further physical constraints.
In the first place, it must preserve the 
positivity of any initial density matrix:
only in this case, the time-evolution turns out to be consistent with
the interpretation of the state-eigenvalues as probabilities.
In the second place, one has also to care of possible couplings with
another system, not subjected to noise, and therefore to guarantee the positivity-preserving 
character also of the semigroup of maps of the form $\Gamma_t=\gamma_t\otimes U_t$,
as studied in the previous section; 
this is only assured in all possible situations if $\gamma_t$ is completely positive.

Nevertheless, in many applications, 
complete positivity is often dismissed as unnecessary.
This attitude is based on the fact that this property
guarantees full physical consistency of the reduced dynamics
at the price of order relations among 
the parameters describing dissipation ({\it cf.} (\ref{6})), which
may appear to have little physical motivations. 
Further, most phenomenological derivation of reduced dissipative
dynamics lead to semigroup of linear transformation that are
not completely positive. 

Unfortunately, such derivations often
lead to time evolutions that are not even positive.
To avoid inconsistencies, one usually restricts the possible initial states 
to those for which $\gamma_t$ remains positive 
(the so-called ``slippage of initial conditions''). 
This prescription works also in the case of the evolution $\Gamma_t$ for two subsystems, 
provided the initial state is in separable form:
$\rho(0)=\sum_i p_i\, \rho^{(1)}_i\otimes\rho^{(2)}_i$,
$p_i\geq0$, $\sum_i p_i=1$, where $\rho^{(1)}_i$ and $\rho^{(2)}_i$
are admissible states for the first and second subsystems, respectively.

On the contrary, as shown in the previous section, when the initial state 
$\rho(0)$ is not in factorized form, 
the evolved matrix $\rho(t)=\Gamma_t[\rho(0)]$ fails to be positive for all times.
In keeping with the same attitude adopted for a single subsystem dynamics
$\gamma_t$, to cure this additional inconsistency 
one has to further restrict the domain of applicability of $\Gamma_t$. 
However, this is again a temporary solution; indeed, the whole discussion needs 
be repeated when three or more subsystems are considered.

These considerations seems to suggest that there is an intrinsic
incompatibility between the existence of entangled states
and the slippage prescription adopted to cure the
inconsistencies that non-completely positive,
or even non-positive, reduced dynamics might produce.
Note that this conclusion can not be dismissed as being purely academic;
on the contrary, it seems to have a direct experimental relevance:
as mentioned before, the scenario studied in the previous
sections might have an actual realization in quantum optics.
From this perspective, the widely used cure of redefining 
the initial conditions in case of non-positive Markovian dynamics 
does not appear to be completely satisfactory.

\end{document}